\documentclass[aps,prd,twocolumn,groupedaddress,nofootinbib]{revtex4-1}

\usepackage{graphicx,color}
\usepackage{latexsym,amsmath,amssymb,graphicx,booktabs}
\usepackage{hyperref}
\definecolor{MyBlue}{rgb}{0.15,0.15,0.70}
\hypersetup{
colorlinks=true,
citecolor=MyBlue,
linkcolor=MyBlue,
urlcolor=MyBlue
}

\newcommand{\nn}{\nonumber}

\newcommand{\iBox}{\Box^{-1}}

\renewcommand\({\left(}
\renewcommand\){\right)}
\renewcommand\[{\left[}
\renewcommand\]{\right]}

\def\lsim{\raise 0.4ex\hbox{$<$}\kern -0.8em\lower 0.62
ex\hbox{$\sim$}}

\def\gsim{\raise 0.4ex\hbox{$>$}\kern -0.7em\lower 0.62
ex\hbox{$\sim$}}

\def\lbar{{\hbox{$\lambda$}\kern -0.7em\raise 0.6ex
\hbox{$-$}}}

\newcommand\eq[1]{eq.~(\ref{#1})}

\newcommand\pa{\partial}
\newcommand\p{\partial}

\newcommand\ee{\end{equation}}
\newcommand\be{\begin{equation}}
\def\bea{\begin{array}}
\def\eea{\end{array}}\def\ea{\end{array}}
\newcommand\ees{\end{eqnarray}}
\newcommand\bees{\begin{eqnarray}}
\def\nn{\nonumber}

\def\dslash{\hspace{-1mm}\not{\hbox{\kern-2pt $\partial$}}}
\def\Dslash{\not{\hbox{\kern-4pt $D$}}}
\def\pslash{\not{\hbox{\kern-2.1pt $p$}}}
\def\kslash{\not{\hbox{\kern-2.3pt $k$}}}
\def\qslash{\not{\hbox{\kern-2.3pt $q$}}}

\def\p1{{\bf p}_1}
\def\p2{{\bf p}_2}
\def\k1{{\bf k}_1}
\def\k2{{\bf k}_2}

\newcommand{\emn}{\eta_{\mu\nu}}

\newcommand{\gmn}{g_{\mu\nu}}

\newcommand{\gMN}{g^{\mu\nu}}

\newcommand{\pam}{\pa_{\mu}}

\newcommand{\pan}{\pa_{\nu}}

\newcommand{\Gmn}{G_{\mu\nu}}

\newcommand{\Tmn}{T_{\mu\nu}}

\newcommand{\dddM}{\kern 0.2em \raise 1.9ex\hbox{$...$}\kern -1.0em \hbox{$M$}}
\newcommand{\dddQ}{\kern 0.2em \raise 1.9ex\hbox{$...$}\kern -1.0em \hbox{$Q$}}
\newcommand{\dddI}{\kern 0.2em \raise 1.9ex\hbox{$...$}\kern -1.0em\hbox{$I$}}
\newcommand{\dddJ}{\kern 0.2em \raise 1.9ex\hbox{$...$}\kern-1.0em
\hbox{$J$}}
\newcommand{\dddcalJ}{\kern 0.2em \raise 1.9ex\hbox{$...$}\kern-1.0em
\hbox{${\cal J}$}}

\newcommand{\dddO}{\kern 0.2em \raise 1.9ex\hbox{$...$}\kern -1.0em
\hbox{${\cal O}$}}
\def\dddz{\raise 1.5ex\hbox{$...$}\kern -0.8em \hbox{$z$}}
\def\dddd{\raise 1.8ex\hbox{$...$}\kern -0.8em \hbox{$d$}}
\def\dddbd{\raise 1.8ex\hbox{$...$}\kern -0.8em \hbox{${\bf d}$}}
\def\ddbd{\raise 1.8ex\hbox{$..$}\kern -0.8em \hbox{${\bf d}$}}
\def\dddx{\raise 1.6ex\hbox{$...$}\kern -0.8em \hbox{$x$}}

\newcommand{\mplr}{m_{\rm Pl}}

\begin{document}


\title{Perturbative loop corrections and nonlocal gravity}


\author{Michele Maggiore}
\affiliation{D\'epartement de Physique Th\'eorique and Center for Astroparticle Physics, Universit\'e de Gen\`eve, 24 quai Ansermet, CH--1211 Gen\`eve 4, Switzerland}



\begin{abstract}
Nonlocal  gravity has been shown to provide a phenomenologically viable infrared modification of general relativity. A natural question is whether the required nonlocality can emerge from perturbative quantum loop corrections due to light particles. We show that  this is not the case. For the value of the mass scale of the non-local models required by cosmology, the perturbative form factors obtained from the loop corrections, in the present cosmological epoch, are in the regime where they are local. The mechanism behind the generation of the required nonlocality must be more complex, possibly related to strong infrared effects and non-perturbative mass generation for the conformal mode.
\end{abstract}


\maketitle

{\em 1. Introduction.} In the last few years, together with various collaborators, we have proposed and developed a class of non-local infrared (IR) modifications of general relativity, which  appear to have quite interesting cosmological consequences. The first successful model of this type was proposed 
in \cite{Maggiore:2013mea} (see also \cite{ArkaniHamed:2002fu,Dvali:2006su,Dvali:2007kt,Porrati:2002cp,Jaccard:2013gla} for earlier related ideas), and is defined by
the non-local equation of motion
\be\label{RT}
\Gmn -\frac{m^2}{3}\(\gmn\iBox R\)^{\rm T}=8\pi G\,\Tmn\, ,
\ee
where the superscript T denotes the operation of taking the transverse part of a tensor (which is itself a non-local operation). The mass $m$ is a free parameter of the model, which replaces the cosmological constant in $\Lambda$CDM.
A closed form of the action of this model is not currently known. A related model, defined at the level of the action, was introduced in 
\cite{Maggiore:2014sia}, and is defined by 
\be\label{RR}
S_{\rm RR}=\frac{\mplr^2}{2}\int d^{4}x \sqrt{-g}\, 
\[R-\frac{m^2}{6} R\frac{1}{\Box^2} R\]\, ,
\ee
where $\mplr^2=1/(8\pi G)$. Both models, that we referred to as the RT and RR model, respectively, have a viable background evolution at the cosmological level displaying self-acceleration, i.e. the nonlocal term behaves as  an effective  dark energy density~\cite{Maggiore:2013mea,Maggiore:2014sia,Foffa:2013vma}. 
Their cosmological perturbations are well behaved  \cite{Dirian:2014ara,Cusin:2015rex} and fit well
CMB, supernovae, BAO and structure formation 
data~\cite{Nesseris:2014mea,Dirian:2014ara,Barreira:2014kra}. The cosmological perturbations have then been implemented in a Boltzmann code in \cite{Dirian:2014bma,Dirian:2016puz}. This allowed us to perform Bayesian parameter estimation and a detailed quantitative comparison with $\Lambda$CDM. The result is that the RT model (\ref{RT})  fits the data at a level which is statistically indistinguishable from  $\Lambda$CDM. In contrast, using the Planck~2015 data and an extended set of BAO observations, we found in \cite{Dirian:2016puz} that the RR model (\ref{RR}),  even if by itself  fits the data at a fully acceptable level, in a Bayesian model comparison with $\Lambda$CDM or with the RT model is  disfavored. The RT model can be considered as a nonlinear extension of the RR model, since the two models become the same when linearized over Minkowski space, so we expect that its action would contain further non-linear terms with respect to the simpler action of the RR model. Since the  observational data point toward the importance of these nonlinear term, in \cite{Cusin:2016nzi} we have explored some other non-linear extension of the action (\ref{RR}), suggested by conformal symmetry. In particular, we found that the model defined by the action
\be\label{6RR}
S_{\rm cRR}=\frac{\mplr^2}{2}\int d^4x\, \sqrt{-g}\, 
\[ R-\frac{m^2}{6} R\frac{1}{(-\Box+  \frac{1}{6}R)^2} R\]\, .
\ee
appears to work quite well [see also \cite{Mitsou:2015yfa} for a study with the more general operator 
$(-\Box+\xi R)^{-2}$]. Even if a full analysis of its cosmological perturbations has not yet been performed, from the equation of state of the effective dark energy we expect that its predictions will deviate from that of 
$\Lambda$CDM less than the predictions  of the RT model (which in turn is closer to $\Lambda$CDM than the RR model), and therefore will be consistent with the data (and possibly difficult to distinguish from $\Lambda$CDM).

Another interesting aspect of these models is that they can be nicely connected with the Starobinsky inflationary model, providing a simple  model that describes both inflation in the early Universe and dark energy at late times. A unified model of this type has been first proposed in \cite{Maggiore:2015rma}
(see also \cite{Codello:2015pga}\cite{Cusin:2016nzi}\cite{Codello:2016neo}), where we suggested to unify the model (\ref{RR}) with the Starobinsky model, through an action of the form
\be\label{6RRStar2}
S=\frac{\mplr^2}{2}\int d^4x\, \sqrt{-g}\, 
\[ R+\frac{1}{6M_{\rm S}^2} R\( 1- \frac{\Lambda_{\rm S}^4}{\Box^2}\) R\]\, ,
\ee
where $M_{\rm S}\simeq 10^{13}$~GeV is the mass scale of the Starobinski model and $\Lambda_{\rm S}^4=M_{\rm S}^2m^2$. The same can of course be done also for the  model (\ref{6RR}), considering the action \cite{Cusin:2016nzi}
\be\label{6RRStar}
S=\frac{\mplr^2}{2}\int d^4x\, \sqrt{-g}\, 
\[ R+\frac{1}{6M_{\rm S}^2} R\( 1- \frac{\Lambda_{\rm S}^4}{(-\Box+  \frac{1}{6}R)^2}\) R\]\, ,
\ee
or for the RT model, combining the non-local contribution in \eq{RT} with the contribution to the equations of motion coming from the $R^2$ term in the Starobinski model. As discussed in \cite{Cusin:2016nzi}, at early times the non-local term is irrelevant and we recover the standard inflationary evolution, while at late times the local $R^2$ term becomes irrelevant and we recover the evolution of the non-local models. 
This has also been recently confirmed in \cite{Codello:2016neo}, through the explicit numerical integration of the equations of motion. Further work on these nonlocal models has been presented in
\cite{Modesto:2013jea,Foffa:2013sma,Kehagias:2014sda,Conroy:2014eja,Cusin:2014zoa,Dirian:2014xoa,Barreira:2015fpa,Barreira:2015vra}.

\vspace{1mm}
{\em 2. Perturbative loop corrections.}  Given that these nonlocal models are  phenomenologically successful, the next question is whether nonlocalities of this form can emerge, at an effective level, from a fundamental local QFT. In general, loops of massless or light particles induce nonlocal terms in the quantum effective action, so it is natural to ask  whether such perturbative corrections can generate a nonlocal term such as that in \eq{RR}, or in its non-linear generalizations (\ref{RT}) or (\ref{6RR}). In gravity the one-loop corrections induced by  matter fields indeed produce  nonlocal form factors associated to terms quadratic in the curvature, which have been computed in several classic papers  using  diagrammatic or heat-kernel techniques \cite{'tHooft:1974bx,Barvinsky:1985an, Barvinsky:1987uw,Barvinsky:1990up,Gorbar:2002pw,Gorbar:2003yt} (see also  \cite{Birrell:1982ix,Buchbinder:1992rb,Shapiro:2008sf} for textbooks or reviews). 
The resulting quantum effective action has the general form 
\bees
S&=&\int d^{4}x \sqrt{-g}\, \bigg[\frac{\mplr^2}{2}R - R \,k_R(\Box) R\label{formfact} \\
&&\hspace*{20mm}
-C_{\mu\nu\rho\sigma}k_W(\Box)C^{\mu\nu\rho\sigma}\bigg]\, ,\nn
\ees
where $C_{\mu\nu\rho\sigma}$ is the Weyl tensor, and we used as a basis for the quadratic term $R^2$, $C_{\mu\nu\rho\sigma}C^{\mu\nu\rho\sigma}$ and the Gauss-Bonnet term, that we have not written explicitly.
For massless particles, the  form factors $k_R(\Box)$ and $k_W(\Box)$ only contain logarithmic terms plus finite parts, i.e.  $k_{R,W}(\Box)=c_{R,W}\log (\Box/\mu^2)$,
where $\Box$ is the  generally-covariant d'Alembertian, $\mu$ the renormalization point, and $c_R ,c_W$  are coefficients that depend on the number of matter species and on their spin. The form factors generated by loops of a massive particle with mass $M$  are more complicated. In the UV limit, i.e. at energies or curvatures such that $M^2/\Box$ can be treated as small, the form factors have
an expansion of the general form 
\bees
k_R\(\frac{-\Box}{M^2}\)&=&\alpha\log\(\frac{-\Box}{M^2}\)+\beta \(\frac{M^2}{-\Box}\)
\label{expan}\\
&&+\gamma \(\frac{M^2}{-\Box}\)\log\(\frac{-\Box}{M^2}\)+\delta \(\frac{M^2}{-\Box}\)^2+\ldots \, ,\nn
\ees
[and similarly for $k_W(-\Box/M^2)$], as discussed for instance in \cite{Codello:2015mba} using a covariant generalization of the EFT formalism of \cite{Donoghue:1994dn}. In \cite{Codello:2015pga} it was then observed that the logarithmic term, as well as the term $(M^2/\Box)$, have little effect on the cosmological evolution in the present epoch. This  might leave as a dominant contribution the one due to  $(M^2/\Box)^2$ which,  as we know from \cite{Maggiore:2014sia},  generates  a phase of accelerated expansion in the recent epoch. In \cite{Codello:2015pga} it was then concluded that a non-local model such as (\ref{RR}) emerges naturally from the perturbative loop corrections.

The purpose of this short note is to point out that, unfortunately, this is not the case, and the mechanism that generates these nonlocal cosmological models must be more complicated. The crucial point is that the expansion (\ref{expan}) only holds in the UV limit, where the operator $M^2/\Box$ can be treated as small. In a cosmological context, this means that $M^2/H^2\ll 1$, where $H(t)$ is the Hubble parameter.\footnote{More precisely $\Box^{-1}$ is a nonlocal operator, which depends on the whole past history. However, from the time evolution of the auxiliary fields $U=-\iBox R$ and  $V=H_0^2\Box^{-2}R$  shown for instance in Fig.~1 of
\cite{Dirian:2014ara} we see that, up to the present epoch, the estimate $\Box^{-1}\sim 1/H^2(t)$ is correct, up to a factor at most ${\cal O}(10)$, which will be irrelevant for the considerations below.}

To understand when this condition is satisfied, we  observe that we can rewrite \eq{RR} as
\be\label{RR2}
S_{\rm RR}=\int d^{4}x \sqrt{-g}\, 
\[\frac{1}{2}\mplr^2 R- R\frac{M^4}{\Box^2} R\]\, ,
\ee
where 
\be\label{M4m2}
M^4=\frac{1}{12}\mplr^2m^2\, . 
\ee
To obtain a viable cosmological evolution, with an accelerated expansion in the present epoch, we need $m=O(H_0)$, where $H_0$ is the present value of the Hubble parameter. This result was obtained in \cite{Maggiore:2013mea,Maggiore:2014sia,Foffa:2013vma} from the explicit integration of the equations of motion, but  of course the order of magnitude follows from simple dimensional considerations. The non-local term in \eq{RR} is suppressed, with respect to the Einstein-Hilbert term, by a factor of order 
$(m^2/\Box^2)R$. In FRW, after radiation dominance, $R\sim H^2$ and $1/\Box\sim 1/H^2$, so 
$(m^2/\Box^2)R\sim m^2/H^2$. If we want that the non-local term  becomes comparable to the Einstein-Hilbert term near the present epoch, we therefore need $m\sim H_0$. Setting $m\sim H_0$, 
\eq{M4m2} gives (apart from numerical factors of order one)
\be\label{MH0}
M\simeq (\mplr H_0)^{1/2}\, , 
\ee
which is huge compared to $H_0$. Indeed, numerically \eq{MH0} gives
$M=O(10^{-3})\, {\rm eV}$, while $H_0=O(10^{-33})\, {\rm eV}$. This means that, for such a value of $M$, the UV expansion 
(\ref{expan}) is not valid near the present epoch, where we are rather in the opposite regime, $M^2\gg -\Box$. The UV expansion is only valid for  $  M^2/H^2(t)\ll 1 $ which, for the value of $M$ given by \eq{MH0}, in terms of redshift means
$z\gg 10^{15}$. The expansion (\ref{expan}) is therefore meaningful only in the very early Universe. 

In the opposite (IR) limit $M^2\gg-\Box$, a particle with mass $M$ is actually heavy compared to the relevant curvature scale and it decouples, leaving only a local contribution.
As an explicit example, for a massive scalar field  with action
\be
S_s=\frac{1}{2} \int d^4x\, g^{1/2}\, \(\gMN\pam\phi\pan\phi+M^2\phi^2+\xi R\phi^2\)\, 
\ee
the form factors $k_R(-\Box/M^2)$ and $k_W(-\Box/M^2)$ in \eq{formfact} were  computed in \cite{Gorbar:2002pw,Gorbar:2003yt}
in closed form, for $-\Box/M^2$ generic. The result is 
\bees
k_W(\Box) &=& \frac{8A}{15\,a^4}
\,+\,\frac{2}{45\,a^2}\,+\,\frac{1}{150}\,, \label{kR}\\
k_R(\Box) &=&
\bar{\xi}^2A
+ \(\frac{2A}{3a^2}-\frac{A}{6}+ \frac{1}{18}\)\bar{\xi}\\
&&+ A\( \frac{1}{9a^4}- \frac{1}{18a^2}
+ \frac{1}{144} \)
+ \frac{1}{108\,a^2} -\frac{7}{2160}
\,, \label{kW}\nn
\ees
where $\bar{\xi}=\xi -(1/6)$ and
\be
A\,=\,1-\frac{1}{a}\log\,\Big(\frac{2+a}{2-a}\Big)\,, \qquad a^2 =
\frac{4\Box}{\Box - 4M^2}\, .
\ee
In the UV limit one recovers the expansion (\ref{expan}). However, in the opposite limit
$-\Box/M^2\ll 1$ one finds
\be\label{dec}
k_W(\Box),k_R(\Box)={\cal O}(\Box/M^2)\, .
\ee
Therefore in this limit the form factor is local, and small,
corresponding to the decoupling of particles with mass large compared to the momentum scale, which is explicit in the mass-dependent subtraction scheme used in \cite{Gorbar:2002pw,Gorbar:2003yt}
(see also the discussion in sect.~2.3.1 of \cite{Maggiore:2015rma}).

In conclusion, a particle with a mass $M\sim 10^{-3}$~eV (such as a neutrino), naively seems to give a contribution  to the term $(M^2/\Box)^2$ in
\eq{expan}, of the right order of magnitude for reproducing the model (\ref{RR}) with $m\sim H_0$, as required by cosmology. However, for such a particle the expansion (\ref{expan}) is invalid near the present epoch. A neutrino is actually an extremely heavy particle compared to the scale $H_0$, and today it gives a {\em local} contribution of the form (\ref{dec}), suppressed by a factor ${\cal O}(\Box/M^2)\ll 1$. A non-local contributions  proportional to $M^4/\Box^2$ at the present epoch  could only be obtained from hypothetical massive particles with  a mass $M\,\lsim\, H_0\sim O(10^{-33})\, {\rm eV}$. However, according to \eq{M4m2}, this would produce a nonlocal term in \eq{RR} with a totally negligible value  $m\sim H_0^2/\mplr$, rather than $m\sim H_0$.

It should also be observed that, at the level of terms quadratic in the curvature, logarithmic corrections involving graviton loops are not even well defined, since they depend on the gauge used,  and one can even find gauges in which the corresponding  divergences are absent, so that the theory is one-loop finite even off-shell~\cite{Kallosh:1978wt}. A related issue is that the  particle creation due to these terms is a pure quantum noise, and the real effect of particle production only starts from terms of third order in the curvature~\cite{Vilkovisky:2007ny}.\footnote{For this reason, it is interesting to study also the  cosmological effects induced by nonlocal terms cubic in the curvature.  In \cite{tesiMichele} has indeed been studied the effect of adding to the Einstein-Hilbert action a term  $R^2\Box^{-2}R$, or a term $R(\Box^{-1}R)^2$. In both cases, however, no viable cosmological model emerges, already at the background level, see Sect.~4 of \cite{tesiMichele}.}

\vspace{1mm}
{\em 3. Strong-coupling effects in the IR?} The above considerations stimulated us in 
\cite{Maggiore:2015rma} to look for less obvious mechanisms for the generation of the required non-localities. A possibility which is rather intriguing is that the scale $M$ that appears in \eq{RR2}, rather than being identified with the mass of a particle running in quantum loops (which, as we have seen, is not a viable possibility) is actually generated dynamically by strong coupling effects, much as $\Lambda_{\rm QCD}$ in QCD. To stress this different interpretation, in \cite{Maggiore:2015rma} we have indeed denoted the mass scale $M$ as
$\Lambda_{\rm RR}$.
The idea of strong coupling effects in gravity in the far infrared might seem difficult to implement. However, as discussed in  \cite{Maggiore:2015rma}, one can imagine  mechanisms that leads to strong IR effects in GR.  One possibility is related to the running of the coupling constant associated to the $R^2$ term. If the running is such that the coupling is asymptotically free in the UV and grows in the IR, a strong coupling regime could be reached at cosmological distances.

Another interesting possibility, again discussed in \cite{Maggiore:2015rma}, is that the dynamics of the conformal mode could become strongly coupled at large distances. Indeed,
restricting to the dynamics of  the conformal mode $\sigma$, i.e. writing the metric as 
\be\label{gmn}
\gmn(x)=e^{2\sigma(x)}\emn\, ,
\ee
the quantum loops corrections embodied in the anomaly-induced effective action generate a non-trivial kinetic term for the conformal mode~\cite{Antoniadis:1991fa,Antoniadis:2006wq},
\be\label{Sanom4D}
S_{\rm anom}=-\frac{Q^2}{16\pi^2}\int d^4x\, (\Box\sigma)^2\, ,
\ee
where  $Q$ depends on the number and type of conformal massless fields. Thus the conformal mode, which in classical GR is a constrained field,  acquires a propagator $\propto 1/k^4$  because of quantum effects.  The corresponding  propagator in coordinate space  grows logarithmically,
\be
G(x,x')=-(2Q^2)^{-1}\log[\mu^2 (x-x')^2]\, . 
\ee
This growth of the two-point correlation at large distances could in principle generate strong IR effects.
The situation is quite similar to what happens in two dimensions, where a momentum-space propagator 
$\propto 1/k^2$ again generates a logarithmically growing propagator in coordinate space, often resulting in a rich infrared physics. A classic example is the Berezinsky-Kosterlitz-Thouless (BKT) phase transition where, changing the value of the parameter $Q^2$ in front of the propagator, a system can make a phase transition from an ordered phase
to a disordered phase, with generation of a  mass gap. As discussed in
\cite{Maggiore:2015rma}, a non-local term of the form $R\Box^{-2}R$ (or its non-linear generalizations) indeed describes a mass for the conformal mode. Indeed, in the metric (\ref{gmn}) we have
\be
R=-6\Box\sigma +{\cal O}(\sigma^2)\, ,
\ee
and therefore, upon integration by parts,
\be\label{m2s2}
m^2R\frac{1}{\Box^2} R=36m^2 \sigma^2 +{\cal O}(\sigma^3)\, .
\ee
The  non local term  $m^2R\Box^{-2}R$   therefore  provides a diffeomorphism-invariant way of giving a mass  to the conformal mode. A mechanism of dynamical mass generation for the conformal mode would therefore naturally produce a term $m^2R\Box^{-2}R$, or one of its non-linear generalizations such as those in eqs.~(\ref{RT}) or (\ref{6RR}).

\vspace{5mm}
\noindent
{\bf Acknowledgments.} I thank Giulia Cusin, Stefano Foffa, Michele Mancarella and Ilya Shapiro for useful discussions.
The work of MM is supported by the Fonds National Suisse and  the SwissMap NCCR.

\bibliography{myrefs_massive}

\end{document}